\documentclass[conference]{IEEEtran}
\IEEEoverridecommandlockouts
\usepackage{cite}
\usepackage{amsmath,amssymb,amsfonts}
\usepackage{algorithmic}
\usepackage{graphicx}
\graphicspath{ {./images/} }
\usepackage{textcomp}

\usepackage[utf8]{inputenc}
\usepackage{soul}
\usepackage{color}
\usepackage{xcolor}
\def\BibTeX{{\rm B\kern-.05em{\sc i\kern-.025em b}\kern-.08em
    T\kern-.1667em\lower.7ex\hbox{E}\kern-.125emX}}
\begin{document}

\title{Crowdsharing Wireless Energy Services}

\author{
\IEEEauthorblockN{Abdallah Lakhdari}
\IEEEauthorblockA{\textit{School of Computer Science} \\
\textit{The University of Sydney}\\
Sydney, Australia \\
abdallah.lakhdari@sydney.edu.au}
\and
\IEEEauthorblockN{Amani Abusafia}
\IEEEauthorblockA{\textit{School of Computer Science} \\
\textit{The University of Sydney}\\
Sydney, Australia \\
amani.abusafia@sydney.edu.au}
\and
\IEEEauthorblockN{Athman Bouguettaya}
\IEEEauthorblockA{\textit{School of Computer Science} \\
\textit{The University of Sydney}\\
Sydney, Australia \\
athman.bouguettaya@sydney.edu.au}

}

\maketitle

\begin{abstract}

We propose a novel self-sustained ecosystem for energy sharing in the IoT environment. We leverage energy harvesting, wireless power transfer, and crowdsourcing that facilitate the development of an \textit{energy crowdsharing} framework to charge IoT devices. The {\em ubiquity} of IoT devices coupled with the potential ability for sharing energy provides new and exciting opportunities to crowdsource wireless energy, thus enabling a {\em green} alternative for powering IoT devices anytime and anywhere.  We discuss the crowdsharing of energy services, open challenges, and proposed solutions.

\end{abstract}


\section{Introduction}

The pervasive presence of wireless sensors and embedded systems has enabled physical devices to connect to the Internet (i.e., Internet of Things (IoT)) \cite{atzori2010internet}. IoT devices' capabilities are augmented with sensors, computing resources, and network connectivity. Abstracting the capabilities of IoT devices using the service paradigm provides novel IoT services \cite{bouguettaya2017service}. \emph{IoT services} are defined by their \emph{functional} and \emph{non-functional} properties. An example of \emph{IoT services} is sensing health-related information by a smartwatch. The functionality of the services is sensing bio-signals from the body. The accuracy of signals and their freshness represent the non-functional, known as Quality of Service (QoS), properties  \cite{perera2014survey}. 
 
\textit{Crowdsourcing} IoT services  creates a self-sustained, fast, and inexpensive IoT ecosystem \cite{ren2015exploiting}. People may exchange a wide range of IoT services using the augmented capabilities of their IoT devices. For example, a set of co-located smartphones in a coffee shop may share their idle computing power to any resource-constrained IoT user in the vicinity \cite{habak2015femto}\cite{wu2017revenue}. Another example is a smartphone with low battery power which may elect to receive energy from nearby IoT devices {\em using WiFi} \cite{TranM0B19}\cite{raptis2019online}\cite{lakhdari2020composing}. Our focus is on providing a platform for crowdsourcing energy services, also known as {\em crowdsharing energy}, among IoT devices, anytime and anywhere.

\begin{figure}[!t]
\centering
\includegraphics[width=\linewidth]{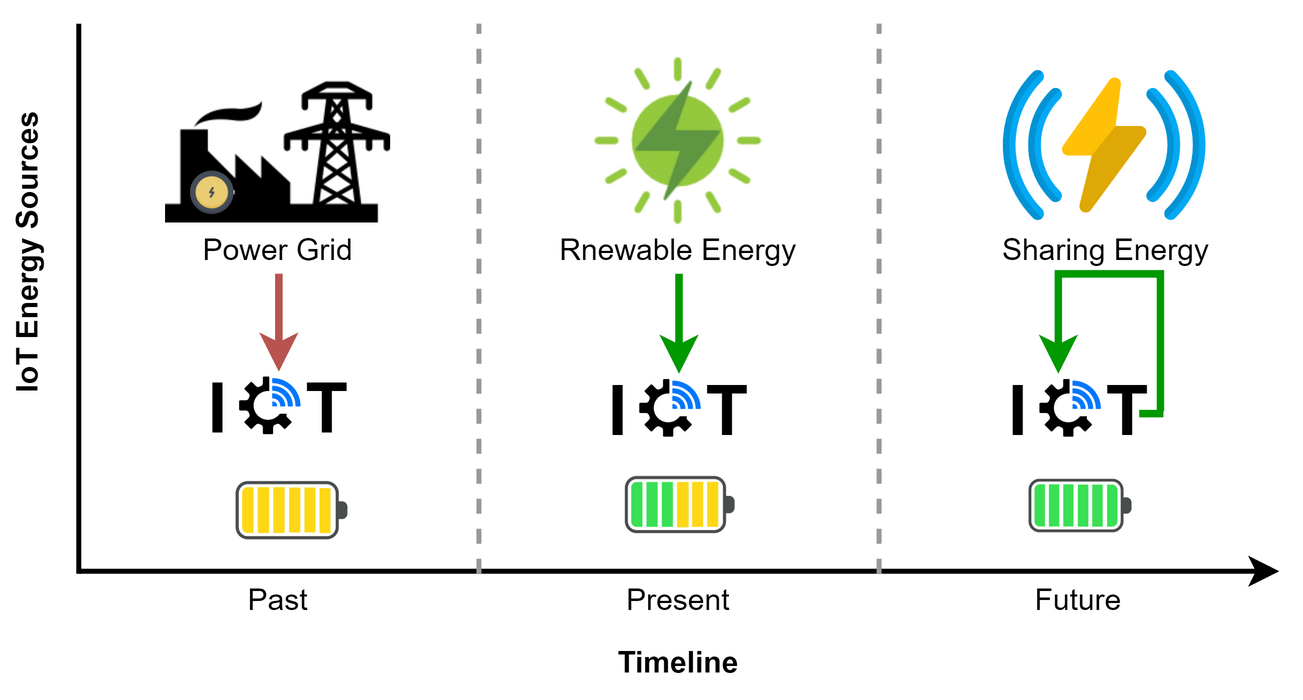}
\caption{IoT Energy Sources Timeline}
\label{timeline}
\end{figure}

Crowdsharing energy presents a novel solution to overcome the increasing demand of energy by IoT devices. Several technologies have been developed over the time to address the energy constraint in the IoT environment (See Fig.\ref{timeline}). Thus far, IoT devices have depended heavily on the power grid \cite{shaikh2015enabling}. Recently, new technologies and solutions have been developed to rely on renewable energy as an alternative resource of energy \cite{shaikh2015enabling}.  We propose the energy crowdsharing {paradigm} as an alternative clean and enhanced source to charge IoT devices.

The energy crowdsharing paradigm leveraging energy harvesting, wireless charging, and crowdsourcing give rise to the concept of crowdsharing energy services. We introduce energy service as the wireless exchange of energy among battery-based IoT devices \cite{lakhdari2018crowdsourcing}. Crowdsharing energy refers to opportunistically crowdsourcing energy services to charge nearby devices. An \emph{energy provider} refers to an IoT device that can share energy. An \emph{energy consumer} is an IoT device that needs energy.  We focus on using \textit{wearables} and IoT devices in the energy crowdsharing paradigm. Wearables refer to IoT devices that can be worn or hand-helds like smart shirts, and smartphones  \cite{seneviratne2017}\cite{gorlatova2015movers}. Wearables have the ability to harvest energy from natural resources, e.g., body heat or kinetic activity \cite{mcfadden_2020}. For instance, Solar Shirt uses flexible solar panels to harvest solar energy and stores it in a battery \cite{reporter_2015}. The harvested energy can be shared with nearby IoT devices as an energy service. Energy services may be achieved with the emergence of new technologies known as ``Over-the-Air wireless charging'' \cite{OvertheAirCharger}. For example, Energous developed a technology to enable wireless charging up to a distance of 4.5 meters\footnote{www.energous.com}. 

\begin{figure*}[!t]
\centering
\includegraphics[width=0.95\textwidth]{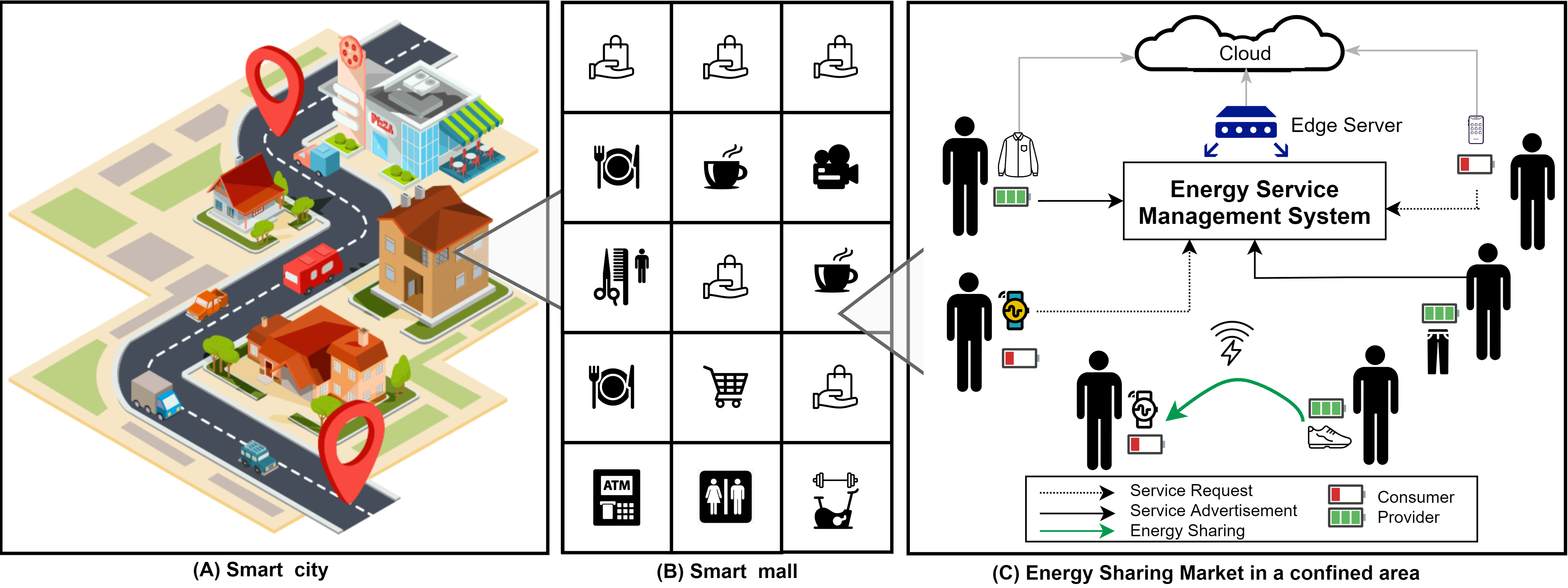}
\caption{Wireless Energy Crowdsharing Market}
\label{Deployment}
\end{figure*}

The energy crowdsharing paradigm in the IoT environment has a twofold impact on both IoT users and the environment. (i) The idea of recycling spare energy promotes a green IoT environment by depending less on the power grid to charge IoT devices. (ii) Enabling wireless charging between IoT devices provides convenience to IoT users and less apprehension to charge their devices. The convenience is a result of allowing IoT users to acquire energy anywhere and anytime from surrounding devices.

We introduce the concept of {\em crowdsourced IoT energy market} where IoT devices exchange energy services. The market could be created by a group of IoT users in a {\em microcell} (See Fig \ref{Deployment}). A microcell is a confined area in a smart city where people may gather such as coffee shops, restaurants, and museums. The prospected crowdsourced IoT energy market is a dynamic environment consisting of providers and consumers that move spatially and temporally across microcells boundaries. The deployment of this crowdsourced market is strongly dependent on the participation of IoT devices. Energy consumers may subscribe to the crowdsharing ecosystem to fulfill their energy requirements. Energy providers, on the other hand, may share their harvested or spare energy {\em altruistically} to contribute to a {\em green} IoT environment. They also may be motivated by  {\em egotistic} purposes through a set of {\em incentives} \cite{abusafia2020incentive}. Due to the limited resources of wearables, a single energy request may not be fulfilled by one energy service \cite{lakhdari2018crowdsourcing}. In such cases, we may need to combine (\textit{i.e. compose}) multiple energy services for one request or vise versa. Additionally, energy service providers and consumers may have different Spatio-temporal preferences. Therefore, these preferences have to be considered in advertising, matching, and composing energy services and requests.

This paper maps out a strategy to leverage the emerging technologies of energy harvesting and wireless power transfer to unlock the full potential of the energy crowdsharing paradigm. We envision the architecture of a green energy ecosystem where IoT devices may exchange energy seamlessly. We identify the major obstacles that hinder the development and potential deployment of the crowdsourced IoT energy market in the real world. We propose and draw a road map for future research directions to achieve this novel vision.\looseness=-1

The remaining of the paper is organized as follows. Section \ref{Benefits} presents the advantages of the energy crowdsharing ecosystem. Section \ref{architecture} presents the architecture of the ecosystem. Section \ref{OpenChallenges} discusses in detail the uprising challenges to implement the envisioned ecosystem. Section \ref{FutureDirections}  highlights the future research directions. Section \ref{conclusion} concludes the paper.

\section{Crowdsharing energy services}
\label{Benefits}

Sharing energy services offer multiple advantages to the environment and IoT users. In what follows, we discuss each of the advantages in detail. 

\begin{itemize}

    \item \textbf{Eco-friendly:}
  The global energy consumption of IoT devices is predicated with a 20\% annual increase to reach 46 TWh by 2025. The massive energy consumption of IoT devices is one of the main contributors to  6\% - 8\% of the global carbon footprint caused by information and communication technologies \cite{shaikh2015enabling}.  Crowdsharing energy contributes to reducing the carbon footprint by creating a self-sustained green IoT environment. The energy services rely on recycling spare energy or harvesting green energy from renewable resources such as body heat, solar, or kinetic activity \cite{gorlatova2015movers}.

    \item \textbf{Spatial freedom:}
Crowdsharing energy services provide spatial independence for IoT users. Energy-constrained IoT devices are regularly required to be tethered to power outlets or carrying power banks. Both charging-cords and charging-pads are the last obstacle for IoT devices to achieve their \textit{spatial freedom}. The prospected crowdsourced IoT environment is expected to enable wireless energy delivery up to five meters. This allows the seamless delivery of energy services. 

    \item \textbf{Flexibility: } 
Crowdshared energy services are typically represented by time intervals. In this environment, consumers and providers may invoke or offer services for the full-time interval or only partially according to their preferences \cite{lakhdari2018crowdsourcing}\cite{abusafia2020incentive}, i.e., they do not have any lock-in contracts like traditional services. In addition, providers and consumers may extend their stay-time to offer or receive an energy service \cite{lakhdari2020Elastic}. These features offer flexible charging services in the crowdsharing environment.
     
    \item \textbf{Energy on demand: }
 The miniaturization of IoT devices restricts the size of their batteries\cite{pasricha2020survey}. The small size of the battery results in a limited energy storage capacity. This limitation restrains the capabilities of the IoT devices. Energy consumption optimization has been extensively researched to enable long-lasting working-time on a single charge. However, there is still a gap for further energy optimizations. Hence, crowdsharing energy presents a simple and yet effective solution to charge IoT devices and extend their working-time \cite{dhungana2020peer}.
 
    \item \textbf{Convenience:}
    Crowdsharing energy services present a convenient alternative to charge IoT devices as it allows them to be charged wirelessly anywhere and anytime. This \textit{ubiquity} of wireless charging facilitates access to energy for IoT users. In addition, the flexibility in advertising and requesting energy services, as previously discussed, emphasizes the convenience for IoT users. IoT users might subscribe to crowdsharing energy without any lock-in contract. They can also move freely while providing or acquiring energy \cite{lakhdari2020fluid}. This convenience promises a wide spread of the crowdsharing ecosystem.
    
    \item \textbf{Growth of the Market:}
    The concept of \textit{wireless crowdcharging} has recently gained increasing interest in industry \cite{bulut2018crowdcharging}. The market for wireless charging has an annual growth rate of 23.4 \% and is predicted to reach \$50 billion dollars in 2027 \cite{WirelessMarket2027}. This recent trend of adopting the wireless charging technology in the industry and the ubiquity of IoT promises the spread and accessibility of crowdsharing energy services \cite{bulut2018crowdcharging}.

\end{itemize}

\section{Proposed crowdsharing ecosystem architecture }
\label{architecture}

The architecture of the system shall include the setup of four pillar components to implement an energy crowdsharing ecosystem. These components comprise the context of the environment,  in-development technology, energy services management system,  and market design. In what follows, we discuss these components in detail.

\begin{itemize}

\item \textbf{Environment:}
We envision a scenario where a number of people gather in different places within the downtown of any major smart city (See Fig.\ref{Deployment}A). The city residents will use their wearables to harvest energy \cite{TranM0B19}. The smart buildings in the city such as a smart mall are split into geographical $microcells$  where a microcell is a confined area such as cafes, movie theaters, etc. (see Fig.\ref{Deployment}B).  There are several IoT devices and wearables in every microcell (see Fig.\ref{Deployment}C). The devices are assumed to be equipped with wireless energy transmitters and receivers. The IoT users might share their spare harvested energy to fulfill the requirements of nearby IoT devices. We identify four components 
in the crowdsharing environment: energy service, energy request, energy provider, and energy consumers. An energy service refers to the wireless transfer of energy among IoT devices. An energy request is an inquiry to receive an amount of energy through wireless means. An \emph{energy provider} refers to an IoT device that may share energy. An \emph{energy consumer} refers to an IoT device that requires energy.   

\item \textbf{Wireless energy transfer technology:}
Two energy technologies are required to implement the crowdsharing ecosystem, namely, \textit{energy harvesting} and \textit{Over-the-Air wireless charging}. Energy harvesters may be any wearable IoT device such as smart clothes, smart shirts, smart glasses, smartwatches, and smartphones. Energy harvesters have the ability to generate energy from natural resources such as body heat, solar, or kinetic activity \cite{gorlatova2015movers}. Existing wearable technology allows for the harvesting of several microwatts to a few watts \cite{khalifa2017harke}.  For instance, Solar Shirt uses flexible solar panels to harvest solar energy and stores it in a battery \cite{reporter_2015}. Another example, PowerWalk kinetic energy harvester produces 10-12 watts on-the-move power \cite{gorlatova2015movers}. PowerWalk harvester can generate enough energy to charge up to four smartphones from an hour walk at a comfortable speed by its user \cite{bionic2019}. Over-the-Air wireless charging technologies are the wireless energy transmitters and receivers that enable the \textit{wireless} transfer of energy\cite{OvertheAirCharger}. For example, ``Energous'' created a technology to enable wireless charging up to a distance of 4.5 meters\footnote{www.energous.com}. IoT devices need to be equipped with both, harvesting and wireless transmitting, energy technologies to be able to crowdshare energy services.

\begin{figure}[!t]
\centering
\includegraphics[width=0.7\linewidth]{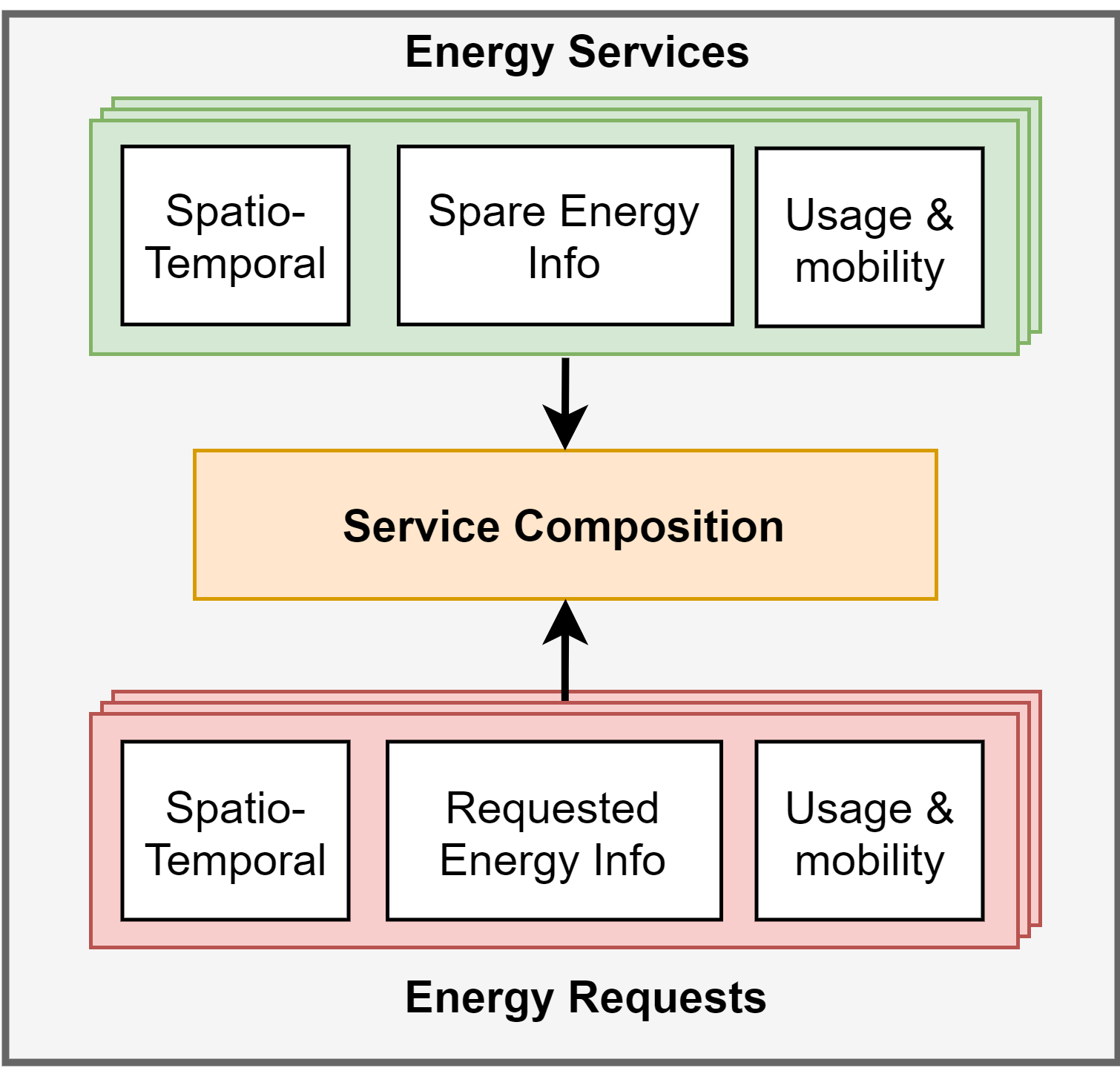}
\caption{ Energy Services Management System}
\label{Arch}
\end{figure}

\begin{figure*}[!t]
\centering
\includegraphics[width=\linewidth]{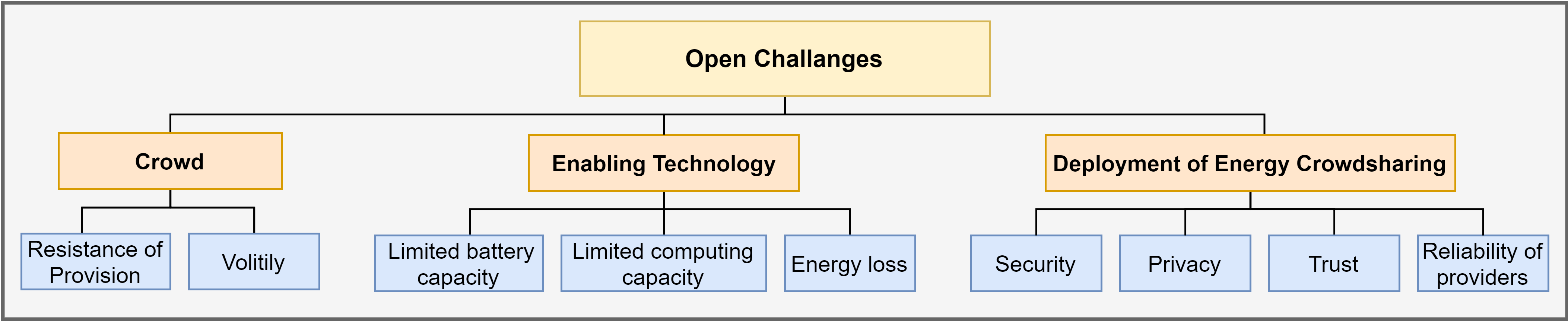}
\caption{ Open Challenges in Energy Crowdsharing Ecosystem}
\label{Open}
\end{figure*}

\item \textbf{Energy services management system} 
The energy service management system is responsible to manage the received energy services and requests. An energy service is defined by spatio-temporal preferences, the amount of available energy, the usage behavior, and the mobility model of the IoT owner (See Fig.\ref{Arch}). Similarly, an energy request is defined by spatio-temporal preferences, the amount of requested energy, the usage behavior, and the mobility model of the IoT owner. A single energy request may not be fulfilled by one energy service due to the limited resources of IoT devices \cite{lakhdari2018crowdsourcing}. In such cases, the system may compose multiple nearby energy services to one request or vise versa.  Additionally, energy service providers and consumers may have different Spatio-temporal preferences. Therefore, the system should consider these preferences in matching and composing energy services and requests. 

\item \textbf{Edge-based crowdsharing market} 
 The crowdsharing market consists of energy providers and consumers in a microcell exchanging energy services (see Fig.\ref{Deployment} C). Typically, an energy provider advertises their services to the energy services management system. Similarly, an energy consumer submits their energy request to the system. All local energy requests and advertisements are handled in the energy services management system at the edge e.g., a router associated with the microcell. Presumably, the management system sets a limit on the amount of requested energy by a consumer and uses a reward system to encourage providers to share energy. Rewards come in the form of stored credits to providers. The collected credits may be used later by the provider to increase the limit on the amount of requested energy when they are in the consumer role. Providers may also be a consumer and vice-versa. Providers receive rewards based on the amount of shared energy.

\end{itemize}

\section{Open challenges}
\label{OpenChallenges}
Unlocking the full potential of sharing energy among IoT devices in a crowdsourced environment requires addressing several challenges. These challenges can be grouped into three major components, the crowd, the enabling technology, and the deployment of the system (See Fig.\ref{Open}). In what follows, we list the encountered challenges for each component. 
\subsection{\textbf{Crowd}}
Sharing energy among IoT devices strongly depends on the crowd. The crowd, in this context, is defined by the owners of the IoT devices. The human factor determines the \textit{willingness} of participation in the energy sharing ecosystem. Typically, the IoT devices owners tend to resist sharing their energy resources \cite{abusafia2020incentive}. Additionally, the dynamic behavior of the crowd may result in volatile energy services. We discuss below the key factors causing the resistance of participation and the volatility of energy services.

\subsubsection{\textbf{Resistance of provision}}
Crowdsourcing energy services create an inexpensive and convenient alternative to charge IoT devices \cite{lakhdari2020composing}. Presumably, people are motivated to participate in the crowdsharing ecosystem as they can easily reach nearby energy sources. However, they might resist sharing their energy \cite{abusafia2020incentive}. Service resistance refers to the unwillingness to provide services due to limited resources. In our context, resistance is defined as the reluctance to provide energy among IoT devices. The resistance of the provider is influenced by the provider's available resources, the service’s required resources, and the current state of the surrounding environment. Therefore, one of the challenges is to predict the resistance of energy providers in offering their energy service and determine a method to overcome their resistance.

\subsubsection{\textbf{Volatility}}

The {volatility} of the energy services, in our context,  refers to the dynamic and unpredictable change in energy services, i.e, change in the energy delivery rate of service. The {volatility} of the energy services is impacted by the following three key aspects. (i) The \textit{availability} of energy services, (ii) the \textit{connectivity} between the consuming and the providing IoT devices, and (iii) the \textit{consistency} of the wireless energy delivery. The aforementioned aspects are influenced by the \textit{mobility} of the crowd, their\textit{ usage behavior} and their \textit{preferences}. {The mobility} of the crowd affects the availability of energy services around an energy consumer. It may also alter the connectivity between the IoT devices. Energy services are provided from IoT devices which are already in use by their owners. Hence, delivering consistent wireless energy from an IoT device to another depends on \textit{the usage} regularity of the device owners. The IoT users' spatio-temporal preferences vary which impacts the availability of the energy services. The impact of each factor is discussed below. . 
\begin{itemize}
 \item {\textbf{Mobility}}:
 IoT devices change their locations frequently according to the movement of their owners. The mobility of the crowd has several effects on the crowdsharing ecosystem. First, the mobility of the crowd, across microcells, determines the availability of energy services and requests.  Second, mobility impacts the connectivity between IoT devices. Since energy services deliver energy through wireless communication channels, they are sensitive to the distance. Therefore, The high mobility of IoT devices within a microcell may result in disconnecting the wireless energy transfer. The disconnection in the delivery off the energy service may result in its failure \cite{abusafia2020Reliability} \cite{lakhdari2020fluid}.  Therefore, managing the mobility of IoT users is a major challenge for providing and consuming energy services in highly dynamic environments. 
 
 \item {\textbf{Usage behavior}}: 
 The providers' energy usage behavior impacts the consistency of delivering energy services \cite{lakhdari2020Elastic}. A highly dynamic usage behavior of the energy providers affects the quality of the shared energy service (QoS). IoT energy service providers may advertise their surplus energy as an energy service while using their devices at the same time. The energy capacity of providing IoT devices changes over time due to the owner's consumption behavior \cite{peltonen2015energy}. Energy services may fluctuate due to the usage of the device owners. Therefore, the energy usage behavior of IoT devices needs to be modeled to ensure a consistent provision of energy services. 

\item {\textbf{Preferences:}} The spatio-temporal features of the IoT energy services and requests are defined based on the preferences of their owners. The preferences are modeled based on the time pattern spent by the IoT owners in regularly visited places, e.g., Workplaces, coffee shops, and food courts. The spatio-temporal preferences of IoT devices are defined using the owners' daily activity model in a smart city \cite{waxman2006coffee}. Matching energy services and requests in a crowdsourced environment is challenging due to the uncertain availability of the services and requests. Predicting the availability of energy services depends on defining the context (i.e., location and time of the day) and the energy consumption behavior of the IoT devices. The contextual information, induced from the daily routines of IoT users, annotates the availability time and location of energy services and requests. The energy consumption behavior determines the amount of the provided or requested energy.  
\end{itemize}

\subsection{\textbf{Enabling Technology:}}

The implementation of the crowdsharing energy ecosystem strongly depends on the recent technology of wireless charging. {Although there is increasing development in the wireless charging technology, there are some limitations in the technology that hinder the full implementation of the ecosystem} \cite{WirelessMarket2027}. Technology limitations include battery capacity, computation capacity, and energy loss.  In what follows, we discuss each of these limitations.

\subsubsection {\textbf{Limited battery capacity}} The design of IoT batteries has been developed to increase the length of their stand-by time. However, it remains a big challenge to keep these devices working while they share energy with other devices. Thus, it is critical to provide an energy-saving technique to increase the IoT devices' working time as long as possible while providing their energy services.

\subsubsection {\textbf{Limited computing capacity}} The limited computing capabilities of the IoT devices present major challenges to share wireless energy. The computing resources are drained by the transfer of energy, the connection establishment, and the conversion of the energy electromagnetic signal between the exchanging devices \cite{na2017energy}. Accordingly, crowdsharing energy algorithms have to consider the required computing resources during wireless energy delivery between devices. 

\subsubsection {\textbf{Energy loss}} Energy services' provided amount may vary from small amounts shared by tiny IoT devices to considerable amounts shared by bigger devices. In the case of small amounts, a consumer may spend more energy to receive the requested energy. The spent energy may include the energy required for service discovery and connection establishment between the consumer and the provider \cite{na2017energy}.

\subsection{\textbf{Deployment of Energy Crowdsharing}}

The number of subscribers to the energy crowdsharing ecosystem is expected to scale largely with the expansion of the IoT devices. This expansion would magnify the threats to energy providers and consumers. It is needed to consider security, privacy, trust, and reliability to achieve the full reach of the crowdsharing ecosystem.

\subsubsection {\textbf{Security}}
Security of IoT devices aims to prevent and protect from  IoT attacks and failure of services \cite{lu2018internet}. The attacks on IoT devices may invade the privacy and confidentiality of the users, infrastructures, data, and devices of the IoT \cite{hassan2019current}. In addition, IoT attacks such as denial-of-service attacks may hinder the provision of services to IoT Users. Securing IoT systems is one of the critical challenges in IoT ecosystems as users may not adopt many IoT systems without a good level of security \cite{liyanage2020iot}. Existing security architectures and protocols are difficult to implement in IoT devices since IoT devices usually have limited resources in terms of computing power and storage size \cite{payton2018envisioning}. In addition, the current architecture of IoT does not effectively meet the security requirements posed by IoT-specific vulnerabilities such as GPS spoofing attacks\cite{deng2016toward}.

\subsubsection {\textbf{Privacy}}
The size of the worldwide digital data created by IoT devices is expected to grow up to 180 Zettabytes by 2020 \cite{kanellos_2016}.  Therefore, the protection of the IoT data and the privacy of the users who generate or consume the data became a major concern in research and industry \cite{mendez2018internet}. Privacy is defined as ``The claim of individuals, groups, or institutions to determine for themselves when, how and to what extent information about them is communicated to others'' \cite{westin1967privacy}. Maintaining privacy in IoT ecosystems may conflict with utilizing IoT data to achieve their functions\cite{sha2018security}. For instance, IoT Data are needed to enhance the energy sharing process by profiling providers and consumers. In the case of energy sharing, the collected IoT data may be considered as personal information such as the IoT users' mobility behavior and their energy consumption model. Therefore, the balance between the privacy and utilization of IoT data is an ongoing challenge.

\subsubsection {\textbf{Trust}}
Trust has been defined as ``The confidence, belief, and expectation regarding the reliability, integrity, ability, and other characteristics of an entity''\cite{yan2014survey}. Trust is required in crowdsourced IoT systems to insure offering high-quality services that will attract users and sustain IoT ecosystems.  In energy sharing,  the trust assessment needs to determine the trust level of the energy provider and the energy consumer.  The trustworthiness of the energy providers may represent an indication of the reputation of the provider, the reliability of their services, and their security level\cite{bahutair2019adaptive}. Energy consumers' trustworthiness may be an indication of their consuming reputation and their security level.  Although trust management has been studied intensively in other domains, it remains a challenge in IoT ecosystems due to several reasons \cite{sha2018security}. First, the dynamic nature of crowdsourced IoT environments makes it challenging to keep an accurate record of the devices' reputations. For instance, IoT devices are usually moving and their existence may not be for long periods. Second, IoT devices usually have no global identity which makes it challenging to keep a globally accessible profile for IoT devices.  Third, the relations among IoT devices are usually temporary which makes it difficult to rely on social relations in evaluating their trust level. Therefore, novel trust management frameworks are needed to determine the reputation of IoT devices \cite{sha2018security}.

\subsubsection {\textbf{Reliability of providers}}
Reliable energy services increase the participation of IoT users in the energy crowdsharing ecosystem \cite{lakhdari2020composing}. The reliability of an energy provider refers to the probability that an IoT energy service will be successfully delivered with the same advertised Quality of Service (QoS) parameters. The reliability of energy service is impacted by the available energy. The source of energy services is an ``in-use'' IoT devices. Therefore, the consistency of the energy transfer may get affected by the usage of the device. The fluctuating behavior of the provider's energy usage may impact the provision of crowdsourced IoT energy services. Hence, it is vital to consider reliability while designing and provisioning energy services. 

\section {Future directions}
 \label{FutureDirections}
This section proposes promising new research directions in crowdsharing. The goal is not to exhaustively list the potential topics, but rather to present some possible directions based on the aforementioned open challenges.

\begin{itemize}
\item \textbf{Incentive Models:}
Several incentive models have been proposed in crowdsourcing environments to increase participation in providing services \cite{capponi2019survey}. However, most of the existing research did not consider the context of crowdsharing energy services \cite{abusafia2020incentive}. Incentives may be used to encourage providers to share their energy. Designing the incentive model shall consider the context of the environment and the behavior of the IoT users. Therefore, designing incentive models is a promising research direction to increase participation in energy provision \cite{dhungana2020peer}.

 \item \textbf{Energy balancing in crowdsharing networks:}
Typically, the initial distribution of energy providers may not be ideal. Some areas might be oversupplied or undersupplied. An area is considered to be undersupplied if the number of energy providers is less than the number of consumers and vice versa. Therefore, the redistribution of energy providers is required to achieve better-balanced energy availability in each area. One possible direction is the use of an incentive model to encourage the migration of providers from oversupplied areas to undersupplied areas.  

\item \textbf{Fairness:} Fairness is a key criterion to efficiently provision energy services and accommodate energy requests. It is challenging to allocate energy and fulfill the requirements of multiple requests when there are limited energy services. Unsatisfied requests would discourage consumers to participate and commit to the crowdshared energy market. Efficient scheduling algorithms present a possible solution to encourage consumers and maximize energy utilization. 

\end{itemize}
\section{Conclusion}
\label{conclusion}

We presented a novel paradigm for energy crowdsharing to enable IoT devices to exchange wireless energy services. We highlighted the advantages of adopting energy crowdsharing on the environment and IoT users. We then envisioned and designed an architecture to implement the energy crowdsharing ecosystem. Lastly, we discussed the open challenges and recommended a road-map for the required solutions.

\bibliographystyle{IEEEtran}
\bibliography{main}
\end{document}